\documentclass{article}
\usepackage{amsmath}%
\usepackage{amsfonts}%
\usepackage{amssymb}%
\addtocounter{page}{1350}%
\begin{document}

\thispagestyle{plain}

\noindent{{\footnotesize {\textit{Journal of Statistical Physics, Vol. 96,
Nos. 5/6, 1999}} ~\hfill\newline ~}}

\vspace{2.5 cm}

\noindent\textbf{\Large A Remark on the Kramers Problem}

\vspace{0.5cm}

\noindent{\small  \textbf{Alex A.~Samoletov}}\noindent \footnote{%
\noindent Laboratoire de Physique, Ecole Normale Sup\'{e}rieure de Lyon,
69364 Lyon cedex 07, France; {\ } Permanent address: Institute for Physics
and Technology, National Academy of Sciences of the Ukraine, 83114
Donetsk, Ukraine. E-mail: samolet@kinetic.ac.donetsk.ua}

\vspace{0.7cm}

{\hfill
\begin{minipage}{10.5cm}
\hrule \vspace{3mm} {\footnotesize We present new point of view on the old
problem, the Kramers problem. The passage from the Fokker-Planck equation
to the Smoluchowski equation, including corrections to the Smoluchowski
current, is treated through an asymptotic expansion of the solution of the
stochastic dynamical equations. The case of an extremely weak force of
friction is also discussed. \vspace{3mm} \hrule \vspace{3mm} \noindent
{\bf KEY WORDS:} ~~Kramers problem; ~~~Klyatskin-Novikov
theory;~~stochastic equation;~~current.}
\end{minipage}}

\vspace{0.7cm}

 ~\textbf{1.} ~~Evolution of a physical system can be ordered in
 multi-time-scales. Details of
evolution on short-time-scale do not need for description in a closed form
of a system evolution on long-time-scale and appears on this scale only in
an average form. The prototype of such kind physical systems is
dissipative Brownian motion of a particle in an external potential field.
In this problem, with the exception of extremely short characteristic time
scales of random forces, there are two time scales: (1) time scales of a
particle motion in an external field; (2) time scales of relaxation (rate
of dissipation) of Brownian particle in a media. It is intuitively
absolutely clear that, if the friction force is strong enough (time of
free motion is extremely short), then probability distribution of a
particle velocity to be rapidly relaxed to the Maxwell distribution and on
this background a particle position will be undergoing to slow process of
diffusion. In the following we deal with the consideration of approximate
reduction of the Fokker-Planck equation for phase-space probability
density to the Smoluchowski equation which deals with probability density
of a particle position only. In the opposite case of extremely weak force
of friction we have energy (or action) variable as evident slow
one.\newline

\textbf{~2.} ~The Kramers problem consist in mathematical description of
approximate reductions of the Fokker-Planck equation for dissipative
Brownian motion of a particle in an external field to the two limiting
cases: (1) to the Smoluchowski equation (extremely strong force of
friction) or (2) to equation for probability density of energy (or action)
variable (extremely weak force of friction)$^{(1)}$. These reduction
procedures are the prototypes of all adiabatic elimination procedures or
procedures of separation on slow and fast variables$^{(2)}$.

{\ ~}The Kramers model$^{(1)}$, firstly formulated for kinetics of
chemical reactions, consists of a particle of mass $m$ moving in an
one-dimensional potential field $U(x)$ under influence of a random force
$f(t)$ and a linear friction force with a constant dissipation rate
$\lambda$. The corresponding set of Langevin equations has the following
form
\begin{equation}
\dot x = u, \qquad m\dot u = -U^{\prime}(x) - \lambda m u + f(t),
\end{equation}
where the random force $f(t)$ is generalized Gaussian $\delta$-correlated
stochastic processes (white noise) with the following properties
(including the fluctuation-dissipation relation)
\begin{equation}
\langle f(t) \rangle= 0, \qquad\langle f(t)f(t^{\prime}) \rangle= 2\lambda
m k_{B} T \delta(t-t^{\prime});
\end{equation}
$\langle\cdots\rangle$ denotes average over all realizations of random
force.

{\ ~}The Langevin (1)-(2) dynamics is stochastically equivalent to the
Fokker-Planck equation for the rate of change of probability density $%
P(u,x;t)$ which has the form (e.g.$^{(2)}$)
\begin{equation}
\partial_{t} P(u,x;t) = - u \partial_{x} P + \frac{1}{m} U^{\prime}(x)
\partial_{u} P + \frac{\lambda}{m}\partial_{u} [uP + \frac{k_{B}T}{m}
\partial_{u} P],
\end{equation}
where $~ $$\partial_{t} =\partial/ \partial t,$$~ $$~ $$\partial_{x} =
\partial/ \partial x , $ and $~ $$\partial_{u} = \partial/ \partial u.$

{\ ~}Keeping in mind (1)-(3) we formulate the problem in the following
manner: in the case of extremely strong force of friction beginning with
(3) or equivalently (1)-(2) to derive the approximate reduction to an
equation for the rate of change of probability density $P(x;t)$ in the
form of asymptotic expansion by the parameter $\lambda^{-1} $:
\begin{equation}
\partial_{t} P(x;t) = - \partial_{x} [\lambda^{-1} J_{S} + o(\lambda^{-1})],
\end{equation}
where $J_{S}$ is the Smoluchowski current,
\begin{equation}
J_{S} = -[U^{\prime}(x) P(x;t) + k_{B}T \partial_{x} P(x;t) ]/m
\end{equation}
In other words it is asymptotics of strong force of friction on time scales $%
\lambda t \gg1$.

{\ ~}This problem has long history starting since 1940, the date of
publication of the Kramers famous work$^{(1)}$. For relevant references
including reviews of the problem see refs. 2--5. The first treatment of
the problem has been down in ref. 6 and the first correct solution has
been down in ref. 7 and then in refs. 8--10. The works$^{(10-14)}$ are of
importance
for the following in respect of treatment of the corrections older then $%
\lambda^{-3} $ which break the Fokker-Planck structure of (4). Most
general treatment of the problem has been down in ref. 15.\newline

\textbf{~3.} ~All of the cited works deal with (3) as the input equation
for a solution of the problem. The purpose of this paper is to take notice
of the fact that (1)-(2) are indeed convenient input equations for an
answer to the problem. With respect to solutions of (1)-(2) we use the
method of asymptotic expansion by the parameter $\lambda^{-1}$. In the
way, the Fokker-Planck type equations (4) to be derived from approximate
stochastic dynamical equations in each order of $\lambda^{-1}$. Moreover,
Fokker-Planck equation is an approximate equation and in any case must be
derived from input dynamical equations. Convenient and powerful method of
derivation, in particular, of the Fokker-Planck type equations immediately
from stochastic equations has been initiated by Novikov$^{(16)}$ and then
it has been sufficiently developed by Klyatskin$^{(17)}$. In the following
we use this method systematically. In this connection it should be pointed
out that the Klyatskin-Novikov theory, generally, interprets a stochastic
differential equation in the sense of Stratonovich. However, in the case
under consideration it does not matter. We refer the reader to refs. 16
and 17 for further information.

{\ ~}The probability density $P(x;t)$ can be written in the form$^{(17)}$
\begin{equation*}
P(x;t)=\langle\delta(x-x(t))\rangle,
\end{equation*}
where $x(t)$ is a stochastic process and $\delta(\cdots)$ is $\delta $%
-function. Differentiating this definition by time we obtain the equation
\begin{equation}
\partial_{t}P(x;t)=-\partial_{x}\langle\dot{x}(t)\delta(x-x(t))\rangle
\equiv-\partial_{x}J(x;t),
\end{equation}
which has the form of a conservation law and is the proforma for an
equation of the type (4). If $x(t)$ is defined by (1), our problem is in
calculation of asymptotic expansion of $\dot{x}(t)$ by $\lambda^{-1}$ and
then the corresponding average in (6). Further insight is gained by making
the following construction. Rewrite (1) for $\lambda t\gg1$ in the form
\begin{equation}
\dot{x}(t)=-\frac{1}{m}\Lambda^{-1}U^{\prime}(x)+\frac{1}{m}\xi(t),
\end{equation}
where operator $\Lambda$ has the form $\Lambda={d}/{dt}+\lambda,$ ~~and
the Ornstein-Uhlenbeck process is introduced:
\begin{equation*}
\xi(t)=\Lambda^{-1}f(t)=\exp(-\lambda t)\int_{0}^{t}\exp(\lambda t^{\prime
})f(t^{\prime})dt^{\prime}.
\end{equation*}
Formal expansion of $\Lambda^{-1}$ by $\lambda^{-1}$ has the form
\begin{equation*}
\Lambda^{-1}U^{\prime}(x)=\frac{1}{\lambda}\sum_{n=0}^{N}\frac{(-1)^{n}}{%
\lambda^{n}}\left( \frac{d^{n}}{dt^{n}}\right) U^{\prime}(x)+\cdots.
\end{equation*}
Hence, (7) can be written in the form%
\begin{align*}
\dot{x}(t) & \sim\left[ -\frac{1}{m\lambda}U^{\prime}(x)+\frac{1}{m}\xi(t)%
\right] +\frac{1}{m\lambda^{2}}\left[ U^{\prime\prime}(x)\dot
{x}(t)\right]
\\
& \\
& -\frac{1}{m\lambda^{3}}\left[ U^{\prime\prime\prime}(x)(\dot{x}%
(t))^{2}+U^{\prime\prime}(x)\ddot{x}(t)\right] +\cdots.
\end{align*}
The $\dot{x}(t)$, $\ddot{x}(t)$, and so on, can be excluded from right
hand side of last equation repeatedly using iterations of this equation
and its
time  derivatives. Then, and it is important, we are in need of expansion by $%
\lambda^{-1}$ of the stochastic process $\xi(t)$ or, more precisely, of
expansion by $\lambda^{-1}$ of the average in (6) which involves $\xi(t)$.
First of all we must remark that the derivatives $\dot{f}(t)$ and so on,
have sense only as derivatives of the generalized stochastic process $f(t)$$%
^{(18)}$ and break the simple Fokker-Planck structure of (6) as of an
second order partial differential equation. It is evident in the frame of
Klyatskin-Novikov theory$^{(17)}$. Namely in the process of calculation of
the corresponding averages according to ref. 17 we easy detect a complex
form of (6) including memory as well as an integral-operator structure.
Further, in respect of $\lambda^{-1}f(t)$ it is easy to verify$^{(17)}$
that corresponding averages in (6) have factor $\lambda^{-1}$ because the
noise (2) intensity has the order $\lambda$.

{\ ~}Taking into account what has been outlined above the first terms of
expansion of $\dot{x}(t)$ by $\lambda^{-1}$ that lead to the current
$J(x;t)$ expansion up to order $\lambda^{-3}$ (the maximum-order of saving
of the simple Fokker-Planck structure of (6)) can be written in the form
\begin{equation*}
\dot{x}(t)\sim\left( 1+\frac{1}{m\lambda^{2}}U^{\prime\prime}(x)\right) %
\left[ -\frac{1}{m\lambda}U^{\prime}(x)+\frac{1}{m\lambda}f(t)\right]
+\cdots.
\end{equation*}
Substituting last expression into the current $J(x;t)$ (6) and performing
averaging exactly follow Klyatskin-Novikov theory$^{(17)}$ we obtain
\begin{align}
J(x;t) & =\langle\left( 1+\frac{1}{m\lambda^{2}}U^{\prime\prime}(x)\right) %
\left[ -\frac{1}{m\lambda}U^{\prime}(x)+\frac{1}{m\lambda}f(t)\right]
\delta(x-x(t))\rangle+o(\lambda^{-3})  \notag \\
&  \notag \\
& =\left( 1+\frac{1}{m\lambda^{2}}U^{\prime\prime}(x)\right)
J_{S}(x;t)+o(\lambda^{-3}).
\end{align}
where $J_{S}$ is the Smoluchowski current (5). Asymptotic expansions of $%
\dot{x}(t)$ and $J(x;t)$(8) together with (2) lead to conventional
conclusion: the Smoluchowski equation is valid if: {~} $\lambda t\gg1$, {~} $%
l|U^{\prime}(x)|\ll k_{B}T$, {~} $l^{2}|U^{\prime\prime}(x)|\ll k_{B}T$, -{~}%
where a length scale $l=\sqrt{k_{B}T/m\lambda^{2}}$ is introduced. (8)
contains lowest order correction to the Smoluchowski equation$^{(8-10)}$.
Higher order corrections in $\lambda^{-1}$, involving in averaging time
derivatives of the generalized stochastic process $f(t)$, lead to break of
simple structure of the Smoluchowski equation as a second order partial
differential equation. It was pointed out also in traditional approach$%
^{(10-14,5)}$.\newline

\textbf{~4.} ~Consider now the case of extremely weak force of friction.
This case more complicated then previous but not so interesting in
calculation.

{\ ~}Let $m=1$ in (1). In the case of extremely weak force of friction and
on the time-scale $\lambda t\ll1$ the energy $E=u^{2}/2+U(x)$ of
unperturbed system is evident candidate for slow variable. But previously
$E$ must be averaged over period of relatively rapid dynamical
oscillations. It is more convenient, however , to consider the action
variable $J$ instead of $E$, $J=J(E)$.$^{(1)}$ Let $J$ is action variable
averaged over period of rapid dynamical oscillations. Then an equation for
the rate of change of probability density $P(J;t)$ can be written in the
form (see (6))
\begin{equation}
\partial_{t}P(J;t)=-\partial_{J}\langle\dot{J}(t)\delta(J-J(t))\rangle,\quad
P(J;t)=\langle\delta(J-J(t))\rangle,
\end{equation}
In usual way$^{(1)}$ and taking into account the change of time-scale of
the white noise$^{(2,7)}$ we obtain the equations of motion for slow
variables
\begin{equation}
\dot{J}(t)=-\lambda J+\frac{V}{\omega}f(t),\quad\dot{V}=-\lambda\frac {V}{%
\omega}J+f(t),
\end{equation}
where $\omega=\omega(J)=dE/dJ$ is frequency and $V$ is velocity averaged
over period of dynamical motion. Substituting (10) in (9) we obtain
\begin{equation}
\partial_{t}P(J;t)=-\partial_{J}\left[ -\lambda JP+\frac{1}{\omega(J)}%
\langle f(t)V(t)\delta(J-J(t))\rangle\right] .
\end{equation}
For calculation of the average in right hand side of (11) we can use the
Klyatskin-Novikov procedure again. Using (9) and the causality condition$%
^{(17)}$ we obtain for functional derivatives
\begin{equation*}
\frac{\delta V(t)}{\delta f(t)}=1;\quad\frac{\delta J(t)}{\delta f(t)}=\frac{%
V}{\omega}.
\end{equation*}
Taking also into account that $V^{2}/\omega=J$ if $J$=const, according to
ref. 17 we finally obtain
\begin{align*}
\partial_{t}P(J;t) & =\partial_{J}\left[ \lambda JP-\frac{\lambda k_{B}T}{%
\omega(J)}\langle\frac{\delta V(t)}{\delta
f(t)}\delta(J-J(t))-V(t)\partial
_{J}\delta(J-J(t))\frac{\delta J(t)}{\delta f(t)}\rangle\right] \\
& \\
& =\partial_{J}\left[ \lambda JP-\frac{\lambda k_{B}T}{\omega(J)}P+\frac{%
\lambda k_{B}T}{\omega(J)}\partial_{J}\left( JP\right) \right] \\
& \\
& =\partial_{J}\left[ \lambda J+\lambda k_{B}T\frac{J}{\omega(J)}\partial_{J}%
\right] P(J;t).
\end{align*}
It is exactly the Kramers equation. We can derive corrections to this
equation but it is slightly more difficult task then in the case of
extremely strong force of friction and does not take special interest in
the context of this paper.\newline

\textbf{~5.} ~In conclusion, we have presented in a simplest framework a
unique approach to the kinetic equations for slow variables by taking
stochastic dynamical equations as the input instead of the Fokker-Planck
equation. We hope that this approach is general enough.

\vspace{7mm}

\noindent\textbf{ACKNOWLEDGMENTS} \vspace{2mm}

{\ ~}I would like to thank Michel Peyrard for his financial support and
kind hospitality at the ENS-Lyon. This work was also supported in part by
National Foundation for Basic Research (Grant No. F4/310-97), \newpage

\textbf{REFERENCES}

\vspace{5mm}

\noindent%
\begin{minipage}{13.5cm} \footnotesize{
\begin{enumerate}
\item  H. A.~Kramers, \textsl{Physica} \textbf{7}:284 (1940). \item C.
W.~Gardiner, \textsl{Handbook of Stochastic Methods} (2nd ed.) (Springer,
Berlin, 1997). \item P.~H\"anggi, P.~Talkner and M.~Berkovec,
\textsl{Rev.~Mod.~Phys.} \textbf{62}:251 (1990). \item N. G.~van Kampen,
\textsl{Stochastic Processes in Physics and Chemistry}
(North-Holland,\newline\indent Amsterdam, 1984). \item N. G.~van Kampen,
\textsl{Phys.~Rep.} \textbf{124}:69 (1985). \item H. C.~Brinkman,
\textsl{Physica} \textbf{22}:29 (1956). \item R. L.~Stratonovich,
\textsl{Topics in the Theory of Random Noise.~Vol.1.}(Gordon and
Breach,\newline\indent New York, 1963). \item G.~Wilemski,
\textsl{J.~Stat.~Phys.} \textbf{14}:153 (1976). \item U. M.~Titulaer,
\textsl{Physica} \textbf{A91}:321 (1978). \item U. M.~Titulaer,
\textsl{Physica } \textbf{A100}:251 (1980). \item H.~Risken and
H.D.~Vollmer, \textsl{Z.~Phys.} \textbf{B33}:297 (1979). \item H.~Risken
and H.D.~Vollmer, \textsl{Z.~Phys.} \textbf{B35}:177 (1979). \item
H.~Risken, H.D.~Vollmer and H.~Denk, \textsl{Phys.~Lett} \textbf{A78}:22
(1980). \item H.~Risken, H.D.~Vollmer and M.~M\"orsch, \textsl{Z.~Phys.}
\textbf{B40}:343 (1980). \item V. I.~Mel'nikov and S. V.~Meshkov,
\textsl{J.~Chem.~Phys.} \textbf{85}:1018 (1986). \item E. A.~Novikov,
\textsl{ZhETF} \textbf{47}:1919 (1964) [ \textsl{Sov.Phys.-JETP}
\textbf{20}:1990 (1965) ]. See also:
M.~D.~Donsker,~\textsl{Proc.~Conf.~on~the~Theory~and~Applications~of~Analysis~in~Function~Space}
(MIT, Cambridge, 1964).-P.17-30; K.~Furutsu, \textsl{J.~Res.~NBS}
\textbf{67}:303 (1963); V. I.~Klyatskin and V. I.~Tatarskii, \textsl{Teor.
Mat. Fiz.} \textbf{17}:273 (1973). \item V. I.~Klyatskin,
\textit{Statisticheskoe Opisanie Dinamicheskikh Sistem s
Fluktuiruyushchimi Parametrami} (Nauka, Moscow, 1975), in
Russian;~V.~I.~Klyatskin, \textit{Stokhasticheskie Uravneniya i Volny v
Sluchaino-Neodnorodnykh Sredakh} (Nauka, Moscow, 1980), in Russian;
V.~I.~Klyatskin and V. I.~Tatarskii, \textsl{\ Usp. Fiz. Nauk (Sov.Phys.-
Uspekhi) }\textbf{110}:499 (1973). \item I. M. ~Gel'fand and N. Ya.
~Vilenkin, \textit{Generalized Functions.  Vol.4.  Applications
of~Harmonic Analysis} (Acad.~Press, New York, 1964).
\end{enumerate}
}
\end{minipage}

\end{document}